\documentclass{pasj00}
\draft

\begin{document}
\SetRunningHead{Y. Itoh et al.}{}
\Received{2000/12/31}
\Accepted{2001/01/01}

\title{Infrared Search for Young Brown Dwarf Companions around 
Young Stellar Objects in the $\rho$ Ophiucus Molecular Cloud
and the Serpens Molecular Cloud\footnotemark[\ast]}

\footnotetext[$\ast$]{Based on data collected at the Subaru Telescope, 
which is operated by the National Astronomical Observatory of Japan.}
%
\author{Chiaki \textsc{Shirono}\altaffilmark{1},
	Yoichi \textsc{Itoh}\altaffilmark{1}, 
	and
	Yumiko \textsc{Oasa}\altaffilmark{2}
}
\altaffiltext{1}
{Graduate School of Science and Technology, Kobe University, 
1-1 Rokkodai, Nada-ku, Kobe 657-8501}
\email{yitoh@kobe-u.ac.jp}
\altaffiltext{2}
{Faculty of Education, Saitama University,
255 Shimo-Okubo, Sakura, Saitama, Saitama 338-8570} 

\maketitle

\begin{abstract}
We conducted an infrared search for faint companions around 351 young 
stellar objects in the $\rho$ Ophiucus molecular cloud and 
the Serpens molecular cloud.
Nine objects in the Spitzer/IRAC archival images were
identified as young stellar companion candidates.
They showed an intrinsic infrared excess; one object was extremely red both in 
the [3.6] - [4.5] color and in the [4.5] - [5.8] color, and two objects
were red in the [4.5] - [5.8] color.
They were as faint as 15 mag in the [3.6] band.
Follow-up $K$-band spectroscopy revealed that three objects had
deep water absorption bands, indicative of
low effective temperatures.
By comparing the spectra and infrared spectral energy
distributions with synthesized spectra of low-temperature objects,
we derived the effective temperatures and
continuum excess for these objects.
It seems highly likely that one of the three objects 
is a low-mass stellar companion and
two objects are young brown dwarf companions associated with the young
stellar objects.
\end{abstract}

%
%
%
  
\section{Introduction}

A brown dwarf is a sub-stellar object
unable to sustain stable hydrogen burning in its core.
It contracts monotonically
because it releases gravitational energy into radiation.  
Until recently, the faintness of brown dwarfs prevented their
discovery, despite early predictions of their existence (Hayashi \&
Nakano 1963).  
Since Gl 229 B was first identified as
a bona-fide brown dwarf (Nakajima et al. 1995;
Oppenheimer et al. 1995), more than 700 brown
dwarfs have been discovered through photometric and spectroscopic
observations.  
While over
half of stars are found in binary systems, only a few tens of brown dwarfs
have so far been identified as companions to stars.

This deficiency of companion brown dwarfs indicates a distinctive 
brown dwarf formation process.
A brown dwarf may form by the gravitational collapse of a molecular cloud, 
similar to stars (Bonnell \& Bastien 1992; Bate 2000).  
It has also been proposed, however, that a brown dwarf may be born in
a circumstellar disk (Rice et
al. 2003), similar to a suggested giant planet formation mechanism
initiated by the gravitational instability of the disk.
Successive dynamical evolution of such a star-and-brown dwarf system
with other stars in a cluster may eject the brown dwarf
from the system (Reipurth \& Clarke 2001).  
Inutsuka et al. (2010) proposed a massive disk around a protostar.
By gravitational fragmentation, a couple of planetary-mass
objects and brown dwarfs may be formed in the disk.
This "protostar disk" scenario predicts multiple companions. 
In terms of the mass ratio ($q=M_{2}/M_{1}$), 
we expect a large mass ratio for the "cloud" scenario, 
but a small mass ratio for the "circumstellar disk" scenario.
Under the "cloud" scenario, we expect systems with large separations.
Under the "circumstellar disk" scenario, the system separation
should be less than the disk radius unless dynamical ejection.
In short, different formation mechanisms predict different system properties.

Itoh et al. (2005) found DH Tau B, a young brown dwarf companion
to the classical T Tauri star, DH Tau.
Two epoch imaging observations confirmed the common proper motions of 
the primary and the companion.
Spectroscopic observations revealed that
the companion had an effective temperature of 
2700 -- 2800 K.
The mass of DH~Tau~B was estimated to be 30 -- 50 Jupiter mass (M$_{\rm J}$),
based on a comparison of its luminosity and 
effective temperature with evolutionary tracks
(Baraffe et al. 2003; D'Antona et al. 1997).
However, the number of young brown dwarf companions is still very limited.

In this paper, we present the results of an infrared imaging search for
faint companions around young stellar objects (YSOs). 
Because a young brown dwarf is expected to have an effective temperature
as low as 2000 K, a search in infrared wavelengths is efficient.
We used the Spitzer/IRAC archive images and
identified nine objects as companion candidates exhibiting infrared excess.
Among them, two objects have been identified as young brown dwarfs
by near-infrared spectroscopy.

\section{Observations and Data Reduction}

\begin{figure}
\begin{center}
\FigureFile(140mm,270mm){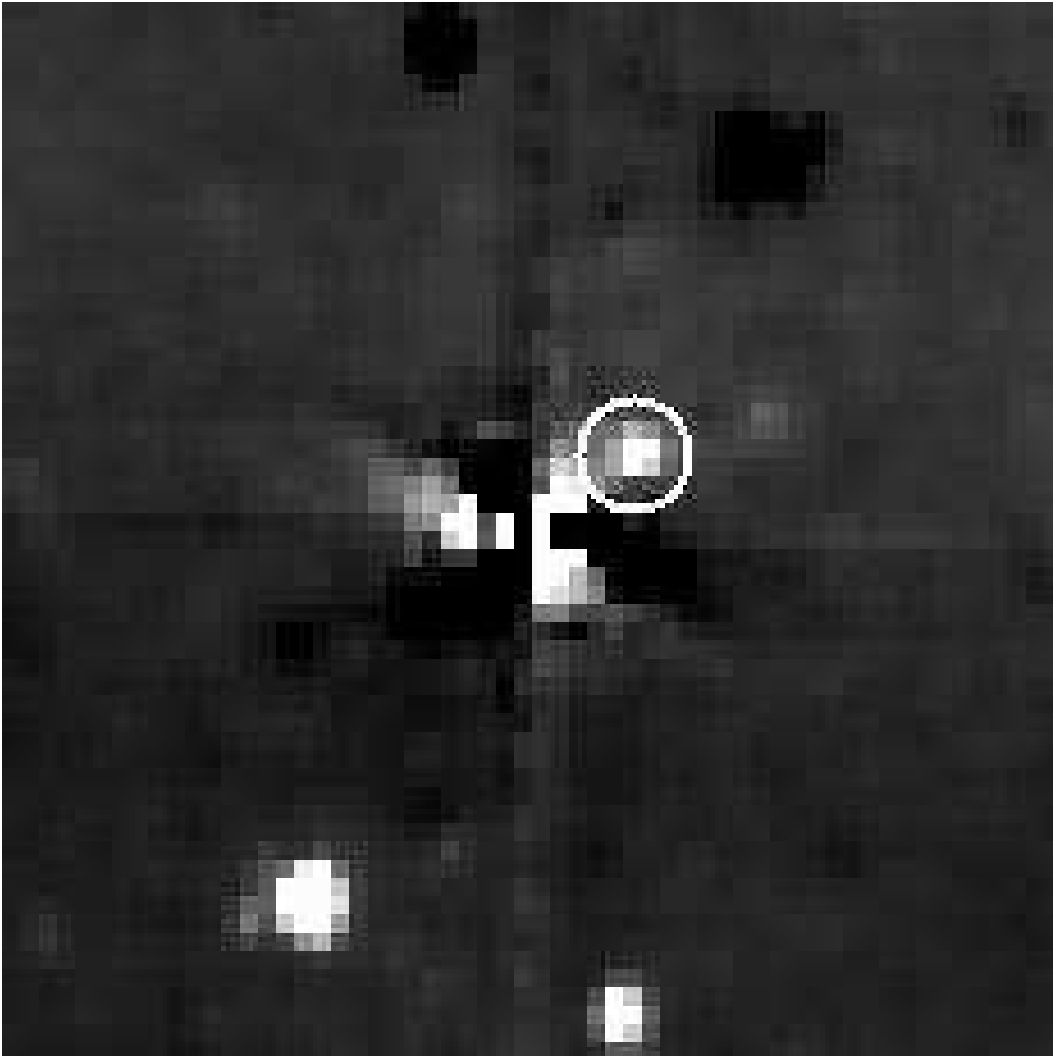}
\end{center}
\caption{An $H$-band image of WL 13.
The central star is located at the center of the image.
The central star halo is suppressed by subtracting
the rotated image of the object itself.
The companion candidate, WL 13 C, is marked by a circle.
The field of view is 70\arcsec $\times $70\arcsec.
North is up, and east is to the left.}
\label{image}
\end{figure}

We searched for faint sources in the vicinity of YSOs in the
$\rho$ Ophiucus molecular cloud and the Serpens molecular cloud. 
The $\rho$ Oph cloud targets were 133 YSOs 
listed in Greene \& Young (1992).
These targets are brighter than 12 mag in the $K$-band. 
We also selected 218 YSOs in the Serpens cloud
from the list in Harvey et al. (2007; hereafter HMH).
We used the Spitzer/IRAC post-basic calibrated data
(PI: Giovanni Fazio for the $\rho$ Oph data and
Neal Evans for the Serpens data).
The IRAC camera imaged the objects in four bands with
a pixel scale of 1\farcs2/pixel.
We first searched for faint objects in the 3.6 $\mu$m image.
A large image was cut into 400$\times$400 pixels centered on the YSO.
We suppressed the bright halo component of the YSO by subtracting a 180\degree~
rotated image of the object frame itself (figure \ref{image}).
Then, the images were visually inspected for faint sources 
originally buried in the halo of the central YSO.
The examined area was circular with a radius of
12\arcsec~ for the $\rho$ Oph data.
For the Serpens data, the radius was 11\arcsec.
In the very close vicinity of the central YSO, the halo component of
the YSO was not
completely removed, which prevented us from detecting faint sources there.
The detection limit was not uniform throughout the data.
It depended on the brightness of the
central star and that of the companion candidates.
We consider that this search is incomplete in the region within 6\arcsec~ from
the central YSO.
We found 16 objects in the examined area across all four bands.
The fluxes of the objects in the 3.6 \micron, 4.5 \micron, and 5.8 \micron~
bands were measured by aperture photometry with the PHOT task in IRAF.
We set the aperture radius as 3 pixels (3\farcs6).
Aperture corrections were applied based on the IRAC data handbook.

For these faint sources, we investigated near-infrared imaging data in the
Subaru archival system (SMOKA).
Two candidates (WL 13 C and VSSG 18 B) 
were previously imaged by the MOIRCS instrument
on the Subaru Telescope.
We retrieved 9 frames in each $JHK$ band.
The exposure times were 120 s for the $J$-band, 90 s for the
$H$-band, and 140 s for the $K$-band.
The FWHM of the PSF was about 0\farcs9, and the 0\farcs46-separated
central stars of WL 13 (WL 13 A and B) were not resolved.
We also imaged one source (HMH 36 B) with the Infrared Camera and Spectrograph
(IRCS) on the Subaru Telescope
on 2008 June 20.
We took five frames for each of the $JHK$ bands.
The exposure time for each frame was 30 s, 10 s, and 5 s for the $J$, $H$, and
$K$ bands, respectively.
Based on the infrared colors, nine objects were identified as companion
candidates, as discussed in Section 3.1.

Near-infrared spectroscopy of three companion candidates 
(WL 13 C, HMH 36 B, and HMH 48 B) was carried out on 
2008 June 20 with IRCS mounted on the Subaru Telescope.  
We used a $K$-band grism. 
The spectral resolution ($R=\lambda/\Delta\lambda)$ was $\sim450$ 
at 2.2~$\mu$m with a $0\farcs6$-width slit.
The seeing size was about $0\farcs5$ and
the adaptive optics system was not used.
The companion candidates were observed at two slit positions with a
2\farcs5 or 5\arcsec~ dithering 
to obtain simultaneous sky data.
We made four or eight exposures at each slit position. 
The integration time of each exposure
was 300 s.
Dwarfs with a spectral type of A0 were used as standard stars 
for spectral calibration.

We used IRAF for spectral data reduction. 
Each dithered pair of object frames was subtracted from each other 
and divided by a dome flat frame. 
We then geometrically transformed the frames to remove the curvature 
of the slit image caused by the grating.  
Wavelengths were calibrated using the sky OH lines.
The wavelength uncertainty was 4 \AA.
Companion candidate spectra were extracted from the transformed images 
using the APALL task.
A one-dimensional spectrum was constructed by integrating 
the spectral image where the intensity of the companion emission was more than 
10 \% of the peak along the slit length  at each wavelength.  
The spectrum was then normalized and combined to produce the final spectrum.  
After the Br$\gamma$ absorption line in a standard star spectrum was
interpolated and removed,
the object spectrum was divided by the standard star spectrum 
and was multiplied by a
blackbody spectrum representing the standard star.
Based on the $K$-band images taken just before the spectroscopic observations,
we confirmed that the contaminated fluxes from the central stars 
in the companion spectra were negligible in all cases.

\section{Results and Discussion}

\subsection{Photometry}

\begin{figure}
\begin{center}
\FigureFile(100mm,100mm){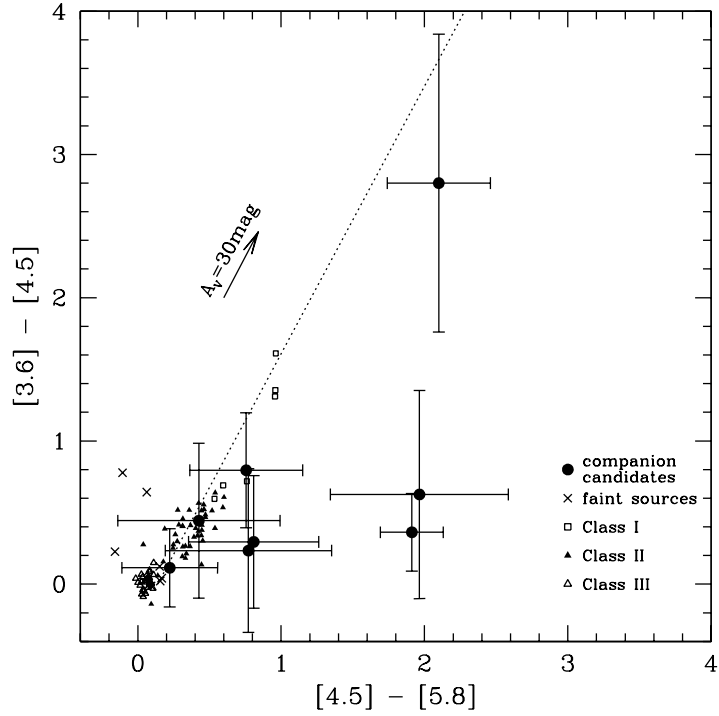}
\end{center}
\caption{IRAC color--color diagram of faint sources.
Companion candidates with infrared excesses are represented by filled circles.
The IRAC colors of Class I objects, Class II objects, and Class III objects
are also shown by open squares, filled triangles, and open triangles,
respectively.
The reddening vector is also shown.
The dotted line starts from the reddest Class III object.
We identified the objects plotted redward this line as the companion candidates.
Crosses represent the IRAC colors of the faint sources
without infrared excesses.}
\label{iraccc}
\end{figure}

Sixteen faint sources were detected around 14 YSOs
out of 351 YSOs surveyed.
But these may be background stars.
We used the IRAC colors to select young low-mass objects associated 
with the clouds.
Figure \ref{iraccc} shows the IRAC color-color diagram of the faint sources.
The IRAC colors of the known YSOs in the Taurus molecular cloud 
(Hartmann et al. 2005) are also plotted.
Class I objects and
the majority of Class II objects are plotted in 
the region redward of Class III objects. 
This red color is caused by radiation from 
circumstellar materials as well as interstellar reddening. 
These two effects have different vectors in the color-color diagram.
The dotted line represents the reddening vector starting from the
reddest Class III object.
The object plotted redward of this line had an
intrinsic excess in the [4.5] - [5.8] color.
More than half the Class II objects and most the Class I objects
exhibited intrinsic excess.
We identified eight objects plotted redward of the line
as young companion candidates.
One object is extremely red both in 
the [3.6] - [4.5] color and in the [4.5] - [5.8] color, and two objects
are red in the [4.5] - [5.8] color.

\begin{figure}
\begin{center}
\FigureFile(100mm,100mm){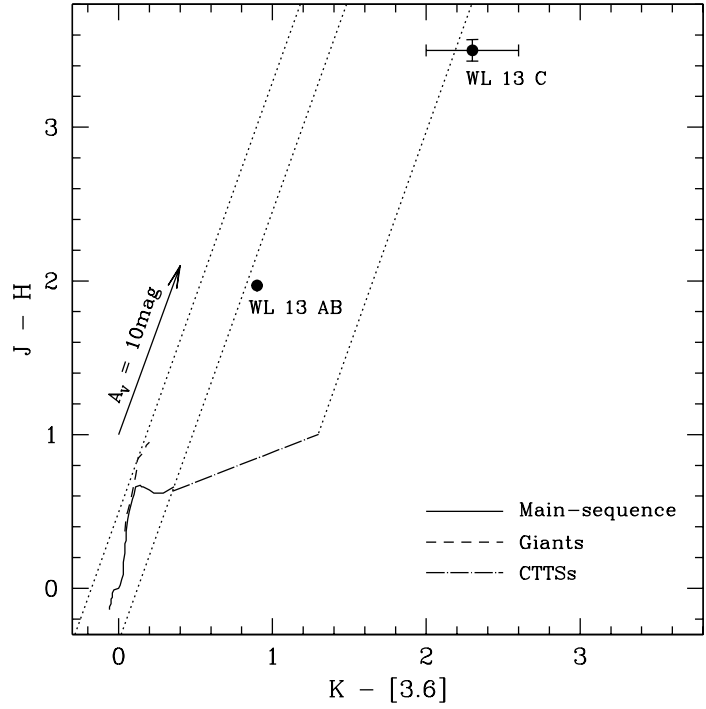}
\end{center}
\caption{$JHK$ and [3.6] bands color--color diagram of WL 13.
The intrinsic colors of main-sequence stars, giants
(Bessell 1988), and classical T Tauri stars
(Meyer et al. 1997) are plotted with the reddening vector.} 
\label{jhkl}
\end{figure}

Among sixteen faint sources, WL 13 C is not plotted in the IRAC color-color
diagram.
We did not obtain its 5.8 \micron~ flux
because of its faintness and the relatively high background.
Instead, we confirmed its infrared excess with the 3.6 \micron~
Spitzer photometry and Subaru's $JHK$ photometries.
Figure \ref{jhkl} shows the $JHK$ and [3.6]-bands color-color diagram. 
The intrinsic colors of main-sequence stars and giants (Bessell \& Brett 1988)
are plotted, 
together with the intrinsic colors of 
classical T Tauri stars (Meyer, Calvet, \& Hillenbrand 1997). 
WL 13 C has an infrared excess, and thus it
is identified as a young companion candidate.

In total, we identified nine young companion candidates.
Most of them are as faint as 14 mag in the 3.6 \micron~ band, which is
$\sim$4 mag fainter than the central YSOs.
The relative positions of the candidates with respect to the central stars,
and the magnitudes of the objects, are summarized in Table
\ref{mag-tab}.
Due to IRAC's large pixel scale,
we did not measure the proper motion of the companion candidates.

\subsection{Near-Infrared Spectra}

\begin{figure}
\begin{center}
\FigureFile(100mm,100mm){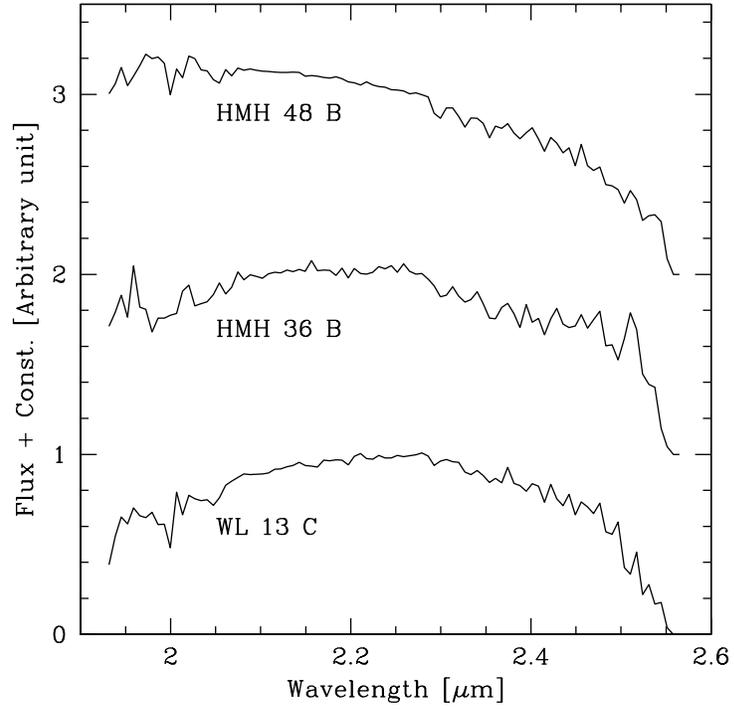}
\end{center}
\caption{Near-infrared spectra of the companion candidates.
Deep absorption bands beyond 2.3 \micron~ indicate low effective temperatures
for the objects.}
\label{sp}
\end{figure}

$K$-band spectra of the three companion candidates are
shown in Figure \ref{sp}.
We found a Br $\gamma$ emission line (2.17 \micron)
from HMH 36 B, with an equivalent width of 2 \AA.
For the other two objects, we did not find Br$\gamma$ lines in emission
or absorption.
All spectra show an absorption band feature at longer wavelengths
($>$2.3\micron),
attributed to H$_{2}$O and CO,
indicating low effective temperatures.
Given this feature and faintness of the objects, we consider that
these objects are low-mass companions to the YSOs.

The strengths of the H$_{2}$O absorption bands are sensitive to
the effective temperature for cool stars (Itoh et al. 2002).
We calculated a reddening-independent index of the H$_{2}$O band
strengths, $Q$, following Wilking et al. (1999). 
With Koornneef's extinction law (Koornneef 1983), the $Q$ index is written as 
\begin{equation}
Q = \left(\frac{F_{1}}{F_{2}}\right)\left(\frac{F_{3}}{F_{2}}\right)^{1.41},
\end{equation}
where $F_{1}$, $F_{2}$, and $F_{3}$ are flux densities in the
2.07 $\micron$ -- 2.13 $\micron$, 2.267 $\micron$ -- 2.285 $\micron$, 
and 2.40 $\micron$ -- 2.45 $\micron$ ranges, respectively.
Late M- or L-type dwarfs have $Q$ values as small as 0.5,
while the $Q$ value is close to unity for objects earlier than K-type.
The $Q$ values of the companions are 
between 0.6 and 0.7, suggesting that they are late M-type stars 
(Table \ref{teff-tab}).

We did not find any other feature, such as Na lines at 2.21 \micron,
previously detected in the spectrum of DH Tau B.
If such metallic lines were detected, we could simultaneously
determine the effective temperature and the surface gravity.

\subsection{Effective Temperatures of the Companions}

We calculated effective temperatures from the
$Q$ index (Oasa et al. in prep.).
The small $Q$ values indicate that the effective temperatures of 
the objects are as low as $\sim$ 3000 K.
However, the H$_{2}$O-band strength also depends on surface gravity.
A low-temperature dwarf and a high-temperature giant have
similar H$_{2}$O absorption band strengths.
Because the surface gravity of a YSO is between that of 
dwarfs and that of giants, 
we derived two effective temperatures, the temperature from
the dwarf scale ($T_{\rm d}$)
and from the giant scale ($T_{\rm g}$), from one $Q$ value.
We expect that a YSO has a temperature between these two temperatures.

\begin{figure}
\begin{center}
\FigureFile(100mm,100mm){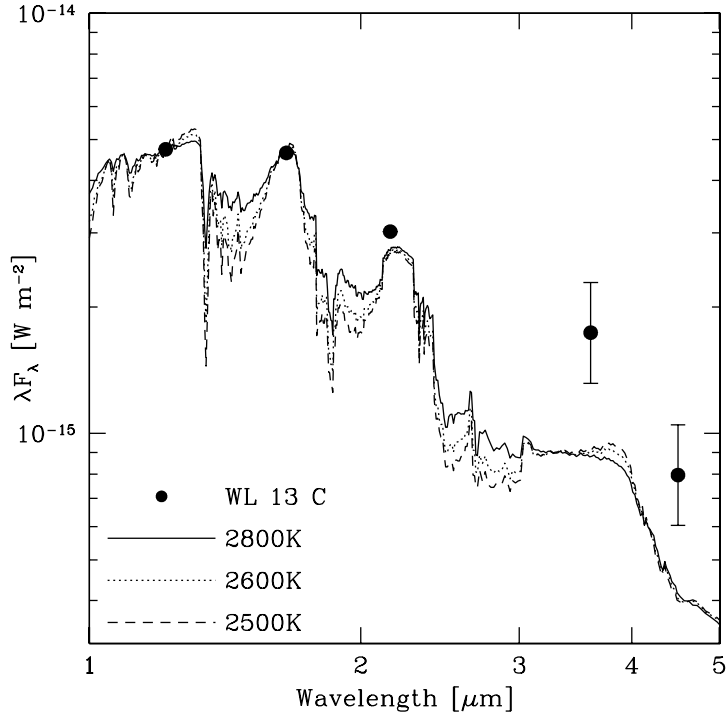}
\end{center}
\caption{
Filled circles show near-infrared fluxes of the companion candidate,
WL 13 C. Interstellar extinction was corrected in each photometric band.
The lines represent synthesized spectra with log($g$) = 5.0.
Discrepancies between the observed fluxes and the synthesized spectra
in longer wavelengths correspond to infrared excess caused by
circumstellar materials.
\label{sed1}
}
\end{figure}

We also considered the continuum excess caused by circumstellar materials.
This could increase the $Q$ index.
The amount of excess is usually measured in the $K$-band
and it is defined as 
$r_{\rm K}=F_{\rm Kex}/F_{\rm K\ast}$,
where $F_{\rm Kex}$ represents the $K$-band flux of the excess emission
and $F_{\rm K\ast}$ is that of the photosphere.
For low-mass YSOs, the $r_{\rm K}$ value decreases with evolution; 
the average $r_{\rm K}$ is 0.6$\pm$0.7 for Class II objects and
0.1$\pm$0.1 for Class III objects (Greene \& Lada 1996).
We estimated the amount of the excess by comparing
the observed $J-K$ colors with the $J-K$ color of the synthesized spectra.
The observed $J-K$ color is sum of the photospheric color,
reddening by interstellar materials, and
continuum excess by circumstellar materials.
The photospheric colors were derived from synthesized spectra of 
Tsuji et al. (2004) 
and Allard et al. (2000) for $T_{\rm d}$ and $T_{\rm g}$, respectively.
The amount of the interstellar reddening was estimated
from the $JHK$ color-color diagram
(WL 13 C; 23 mag, HMH 36 B; 20 mag).
The difference between the reddening corrected $J-K$ color of the object and the
$J-K$ color of the synthesized spectrum corresponds to the continuum excess.
In this procedure, we assumed no excess in the $J$-band.
Because the effective temperature of the companion is as low as 3000 K,
we did not consider any significant $J$-band emission arising from the
circumstellar material of the companion.
Next, the excess was subtracted from the observed spectrum.
We assumed a constant excess in the $K$-band.
Then, the revised $Q$ value ($Q'$) was calculated from the excess-subtracted
spectrum.
The revised temperatures were also estimated.
We iterated this procedure until the newly estimated temperature
was coincident (within 100 K) with the previously estimated temperature
(Table \ref{teff-tab}).
For HMH 48 B, the reddening-corrected $J-K$ color derived from the
broad-band photometry was smaller than the $J-K$ color of the 
synthesized spectrum with the $T_{\rm eff}$ derived from the $Q$ index.
We did not apply the above iteration process to this object.
As a result, the effective temperatures were derived as 2460 K --
3580 K, 2650 K -- 3670 K, and 3140 K -- 4110 K for
WL 13 C, HMH 36 B, and HMH 48 B, respectively.
We present the broad-band photometries and
the fitted synthesized spectra of WL 13 C in Figure \ref{sed1} as
an example.
A large contribution from the continuum excess to 
the $K$-band flux significantly
reduces the $Q$ value, especially on the dwarf scale.
The $r_{\rm K}$ values of the companions are $\sim$0.3 in the
dwarf scale and $\sim$0.0 in the giant scale.
These $r_{\rm K}$ values are consistent with the Class II $r_{\rm K}$ and the
Class III $r_{\rm K}$, respectively.
As described below, the ages of the companions were estimated to be
0.1$\sim$1 Myr on the dwarf scale and more than 10 Myr in the giant scale.
The $r_{\rm K}$ values and the ages of the companions are
consistent with each other in either gravity scale.

\subsection{Masses of the Companions}

We derived the photospheric luminosity of the companions
by comparing the photospheric luminosity of
late-type dwarfs in the $J$-band (Leggett et al. 1996).
This uncertainties in this procedure are described in Itoh et al. (2002).

\begin{figure}
\begin{center}
\FigureFile(100mm,100mm){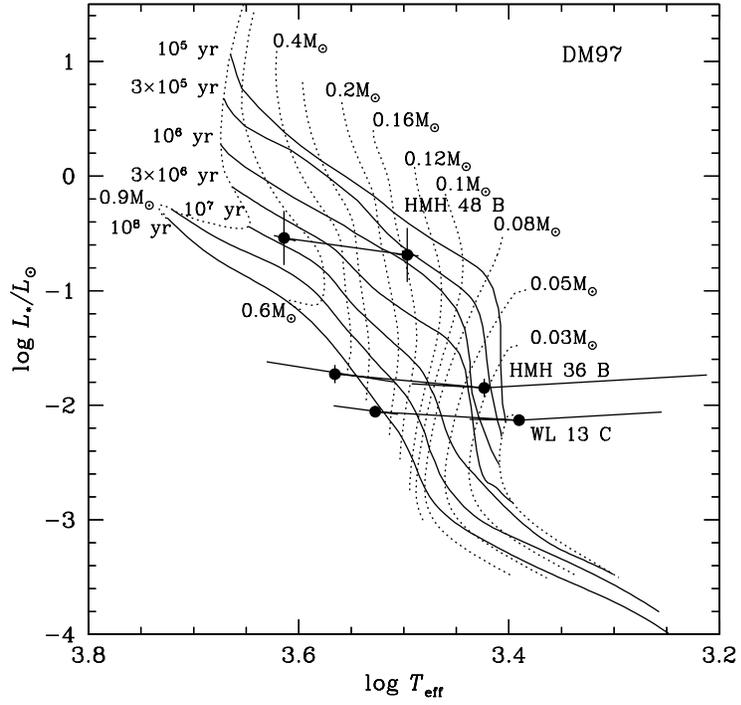}
\end{center}
\caption{HR diagram of the companion candidates with evolutionary tracks from
D'Antona \& Mazzitelli (1997).}
\label{DM98}
\end{figure}

\begin{figure}
\begin{center}
\FigureFile(100mm,100mm){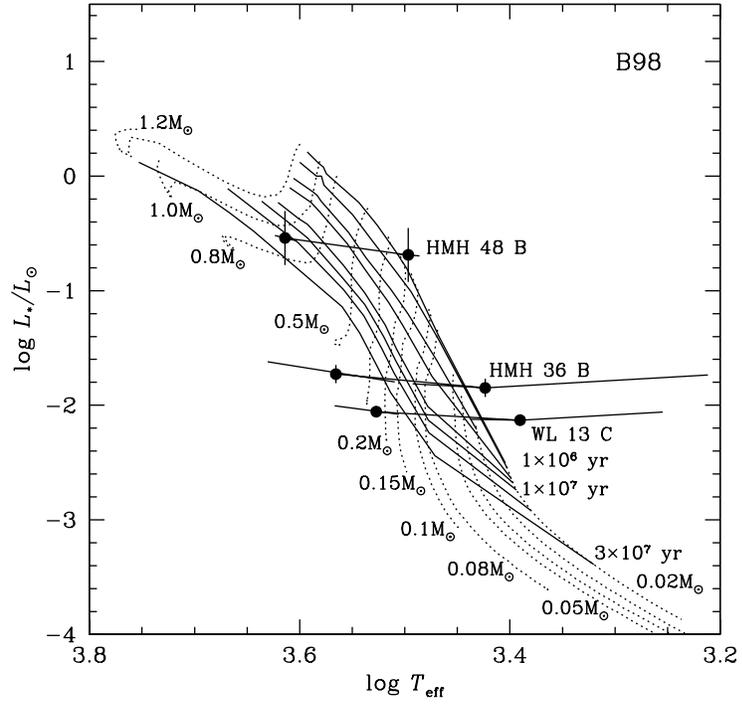}
\end{center}
\caption{HR diagram of the companion candidates with evolutionary tracks of 
Baraffe et al. (1998)}
\label{b98}
\end{figure}

Figures \ref{DM98} and \ref{b98} show the HR diagrams of the companions. 
Evolutionary tracks from D'Antona and Mazzitelli (1997)
and Baraffe et al. (1998) are overlaid.
Each object has two points on the HR diagram.
The right point represents the object position with $T_{\rm eff}$ derived
from the dwarf $Q$ scale, and
thus, it represents the high surface gravity case.
The left point represents the position 
determined from the giant $Q$ scale, and thus, the low surface gravity case.
The masses and ages of the companions were estimated from
the HR diagrams on these two gravity scales (Table \ref{mass-tab}). 
We deduced that WL 13 C and HMH 36 B are young brown dwarfs
and that HMH 48 B is a low-mass star.

However, we noticed that neither the right point nor the left point 
was consistent with the evolutionary tracks in terms of the surface gravity.
The evolutionary tracks indicate that an object has
low-surface gravity in the early phase of evolution (i.e., at the
upper-right in the HR diagram).
We think that the true location of the object must lie somewhere on the line 
between the points.
This idea is supported by DH Tau B (Itoh et al. 2005).
The surface gravity of DH Tau B was derived to be
log($g$)=4.0 or 4.5 (in cgs units) from the equivalent widths
of the metallic lines, as well as the $Q$ index.
This value is between the dwarf surface gravity and the giant surface gravity.
We provide the masses and ages of the companions in Table \ref{mass-tab}
assuming log($g$)=4.0.
We conclude that HMH 48 B is a low-mass YSO and WL 13 B and HMH 36 B are young
brown dwarfs with masses of 30 M$_{\rm J}$ and 80 M$_{\rm J}$.
Precise surface gravities for these objects will be determined from
the equivalent width ratios of metallic lines (Takagi et al. 2010).

WL 13, also called as VSSG 25,
is a classical T Tauri star associated with the
$\rho$ Ophiucus molecular cloud.
It is a binary with a separation of 0\farcs46 (Costa et al. 2000).
The $K$-band flux ratio of the primary to the secondary
is $1.9\pm0.2$.
The spectral type of the binary is M4 and the visual extinction is estimated 
to be 11.7 mag (McClure et al. 2010).
The total mass is derived as 0.4 \MO (Natta et al. 2006).
We found the third body, WL 13 C, at 1400 AU away from the
binary.
Its mass is 0.02 \MO.
Based on the color-color diagram,
the visual extinction of the companion is larger than that of
the central binary (Figure \ref{jhkl}).
The companion may be surrounded by local circumstellar materials.
The mass ratio of the third body to the central binary ($q$) is 0.05,
which is considerably smaller than the average ratios of nearby known binaries.

HMH 36 and HMH 48 are located south of the Serpens Cluster B.
HMH 36 is an object newly discovered by the Spitzer Telescope.
It has a flat spectrum in the Spitzer bands, so that its evolutionary
stage is thought to be between the Class I phase and the Class II phase.
HMH 48 is also an object discovered by Spitzer.
It is classified as a Class III object with $A_{\rm V}$=22.5 mag.

One may consider that the companions are not physically associated with 
the YSOs, but are isolated low-mass YSOs.
From the IMF of the $\rho$ Oph cloud (Marsh et al. 2010),
the expected number of objects located within 10\arcsec~ from
the central star is only 0.012,
if the detection limit is $K = 16$ mag.
We will check the association by proper motion measurements in the next 
few years.

If the objects are physically associated with each other, 
the wide separations of
the brown dwarf companions provides some constraints on the formation mechanism.
If small mass ratio systems are common, 
the "cloud" scenario is not a realistic formation process for brown 
dwarf companions.
In the "circumstellar disk" scenario, such a wide separation for the
brown dwarf companion is 
difficult to achieve without invoking ejection phenomena induced by other objects.
This ejection process
may explain the case for the WL 13 system, because the central object
is a binary system.
Recently, Inutsuka et al. (2010) proposed the
emergence of a massive protoplanetary
disk at the formation stage of a protostar.
Its radius is $\sim$ 100 AU.
This disk is so massive that planetary-mass objects and brown dwarfs are
formed in the disk by gravitational instability.
Because the gravitational fragmentation and formation of planetary-mass
objects and brown dwarfs repeat many times during the main accretion phase
of the central star,
a number of less-massive objects are periodically formed.
Some of these will fall onto the central star, while some remain in the disk
or migrate outward.
If this scenario was an accurate description of 
the WL 13 system and/or the HMH 36 system formation sequence,
then we would expect other low-mass companions around the central star.
Some might be young planetary-mass objects orbiting the star.
Due to the large pixel scale of Spitzer/IRAC,
the search reported here is significantly incomplete, especially
very close to the central stars.
A high-resolution imaging search may reveal large populations of
young brown dwarf companions.

\section{Conclusions}

An infrared search for faint companions around 351 YSOs in the $\rho$ Ophiucus
and Serpens molecular clouds
was conducted with Spitzer/IRAC archived data.
We followed up with near-infrared spectroscopy of three faint companion candidates.
Our main results were as follows:

\begin{enumerate}

\item
Nine objects were identified as the young low-mass companion candidates.
Their infrared colors show an infrared excess.
\item
$K$-band spectra of three objects show a deep absorption feature
from H$_{2}$O and CO.
\item
We derived effective temperatures of 2460 K to 4110 K for the objects
by combining the object spectra with broad-band photometry.
\item
The masses of the objects were estimated.
Two objects are possible young brown dwarf companions with
masses of 30 M$_{\rm J}$ and 80 M$_{\rm J}$.
\end{enumerate}

\bigskip

This work was
supported in part by "The Global COE program:
Foundation of International Center for Planetary Science'' of
the Ministry of Education, Culture,
Sports, Science, and Technology (MEXT).

\clearpage

\begin{table}
\begin{center}
\caption{Magnitudes and Positions of Companion Candidates.}
\label{mag-tab}
\begin{tabular}{ccccccccccc}
\hline
\hline
Object & \multicolumn{5}{c}{Subaru} & \multicolumn{5}{c}{Spitzer} \\
& $J$ & $H$ & $K$ & Sep. & P.A. 
& [3.6] & [4.5] & [5.8] & Sep. & P.A. \\
 & [mag] & [mag] & [mag] & [\arcsec] & [\degree] 
 & [mag] & [mag] & [mag] & [\arcsec] & [\degree] \\
\hline
WL 13 C   & 21.0$\pm$0.0 & 17.5$\pm$0.0 & 15.8$\pm$0.0  &  8.78$^{\dagger}$ & 314.32$^{\dagger}$ 
          & 13.5$\pm$0.3 & 13.3$\pm$0.3 &               & 8.7   & 314.9 \\
VSSG 18 B &              &              & 16.3$\pm$0.0  & 11.39 &  68.39 
          & 15.4$\pm$1.0 & 12.6$\pm$0.3 & 10.5$\pm$0.2  & 11.5  &  66.5 \\
HMH 20 B  &              &              &               &       &        
          & 14.6$\pm$0.3 & 14.3$\pm$0.3 & 13.5$\pm$ 0.3 &  5.73 &  88.2 \\
HMH 36 B  & 19.8$\pm$0.2 & 16.8$\pm$0.1 & 15.1$\pm$0.0  & 12.61 &  28.07 
          & 14.6$\pm$0.2 & 14.4$\pm$0.2 & 14.2$\pm$0.3  & 13.1  &  28.3 \\
HMH 37 B  &              &              &               &       &        
          & 15.8$\pm$0.4 & 15.2$\pm$0.6 & 13.2$\pm$0.2  &  5.5  & 243.4 \\
HMH 40 B  &              &              &               &       &        
          & 14.4$\pm$0.3 & 13.6$\pm$0.3 & 12.8$\pm$0.3  &  6.99 & 211.4 \\
HMH 43 B  &              &              &               &       &        
          & 15.0$\pm$0.4 & 14.5$\pm$0.4 & 14.1$\pm$ 0.4 &  5.27 & 242.3 \\
HMH 48 B  &              & 14.9$\pm$0.6 & 14.1$\pm$0.3  &       &        
          & 14.7$\pm$0.2 & 14.3$\pm$0.2 & 12.4$\pm$0.1  &  6.5  & 273.8 \\
HMH 112 B &              &              &               &       &        
          & 15.0$\pm$0.4 & 14.8$\pm$0.4 & 14.0$\pm$ 0.4 &  7.09 & 123.2 \\
\hline
\hline
\end{tabular}
\end{center}
$^{\dagger}$: The central star is a 0\farcs46 separated binary with 
a $K$-band flux ratio of $1.9\pm0.2$. Due to the poor spatial resolution of the 
observations, the binary was not resolved.
The separation and position angle are relative to the centroid
of the binary flux in the $K$-band.
\end{table}

\begin{table}
\begin{center}
\caption{Effective Temperatures and Photospheric Luminosities of
Companion Candidates.}
\label{teff-tab}
\begin{tabular}{ccccccc}
\hline
\hline
Object & $Q$ & Gravity Scale & $Q'$ & $r'_{\rm K}$ &
Effective Temperature & Luminosity \\
 & & & & & [K] & log($L_{\rm bol}/L_{\odot}$) \\
\hline
WL 13 C  & 0.62$\pm$0.07 & dwarf & 0.53 & 0.38 & 2460 & -2.13 \\
         &               & giant & 0.61 & 0.03 & 3580 & -2.02 \\
HMH 36 B & 0.64$\pm$0.10 & dwarf & 0.58 & 0.24 & 2650 & -1.85 \\
         &               & giant & 0.64 & 0.03 & 3670 & -1.73 \\
HMH 48 B & 0.71$\pm$0.02 & dwarf &      &      & 3140 & -0.69 \\
         &               & giant &      &      & 4110 & -0.54 \\
\hline
\hline
\end{tabular}
\end{center}
\end{table}

\begin{table}
\begin{center}
\caption{Masses and Ages of Companion Candidates.}
\label{mass-tab}
\begin{tabular}{cccccccc}
\hline
\hline
Object & gravity & \multicolumn{2}{c}{DÁntona \& Mazzitelli (1997)} & \multicolumn{2}{c}{Baraffe et al. (1998)} & \multicolumn{2}{c}{Burrows et al. (1997)} \\
 & & Mass & Age & Mass & Age & Mass & Age \\
 & & [$M_{\odot}$] & [Myr] & [$M_{\odot}$] & [Myr] & [$M_{\odot}$] & [Myr] \\
\hline
WL 13 C  & dwarf        & $<$0.017   & 0.1    & $<$0.02    & $<$1      
			& 0.02       & 0.1 \\
         & giant        & 0.2--0.3   & $>$100 & 0.2--0.3   & 100--1000 
			& 0.3        & 100 \\
         & log($g$)=4.0 &            &        & 0.03       & 1-1.7 \\
HMH 36 B & dwarf        & 0.03--0.05 & 0.1--1 & 0.02--0.05 & $<$1      
			& 0.05       & 0.1--1 \\
         & giant        & 0.3--0.5   & $>$100 & 0.3--0.5   & $>$1000   
			& $>$0.3     & $\sim$100 \\
         & log($g$)=4.0 &            &        & 0.08       & 5 \\
HMH 48 B & dwarf        & 0.1--0.2   & 0.1--1 & 0.2        & 1         
			& $<$0.08    & $<$0.1 \\
         & giant        & $>$0.5     & 1--10  & 0.8--1     & 10--100   
			& 0.3        & 0.1--1 \\
         & log($g$)=4.0 &            &        & 0.6        & 5 \\
\hline
\hline
\end{tabular}
\end{center}
\end{table}

\end{document}